# Two-dimensional relative peripheral refractive error induced by Fractal Contact Lenses for myopia control

Manuel Rodriguez-Vallejo[1*], Karina Naydenova [3], Juan A. Monsoriu[2], Vicente Ferrando[3] and Walter D. Furlan[3]

[1]Qvision, Unidad de Oftalmología, Vithas Hospital Virgen del Mar, 04120, Almería

[2]Centro de Tecnologías Físicas, Universitat Politècnica de València, 46022 Valencia, Spain

[3] Departamento de Óptica, Universitat de València, 46100 Burjassot, Spain

[*]Corresponding author: manuelrodriguezid@qvision.es

## 1. ABSTRACT

**Purpose:** To assess the peripheral refraction induced by Fractal Contact Lenses (FCLs) in myopic eyes by means of a two-dimensional Relative Peripheral Refractive Error (RPRE) map.

**Methods:** FCLs prototypes were specially manufactured and characterized. This study involved twenty-six myopic subjects ranging from -0.50 D to -7.00 D. The two-dimensional RPRE was measured with an open-field autorefractor by means of tracking targets distributed in a square grid from -30° nasal to 30° temporal and 15° superior to -15° inferior. Corneal topographies were taken in order to assess correlations between corneal asphericity, lens decentration and RPRE represented in vector components M, $J_0$ and $J_{45}$.

**Results:** The mean power of the FCLs therapeutic zones was 1.32 ± 0.28 D. Significant correlations were found between the corneal asphericity and vector components of the RPRE in the nacked eyes. FCLs were decentered a mean of 0.7 ± 0.19 mm to the temporal cornea. M decreased asymmetrically between nasal and temporal retina after fitting the FCLs with a significant increment of the myopic shift besides 10° (p<0.05) and the induced myopic shift at 25° and 30° decreased with FCLs decentration to temporal cornea. The peak of maximum myopic shift at the peripheral retina (M= -1.3 D) was located at 20°. Two-dimensional maps showed uniform significant differences in extreme positions of the visual field in comparison with the horizontal RPRE for M and $J_0$, but not for $J_{45}$.

**Conclusions:** FCLs measured in myopic eyes showed a similar performance than the reported in previous ray-tracing studies with a small bias explained by the manufacturing process and the lens decentration. Two-dimensional maps are preferable for assessing $J_{45}$ changes due to the lens.

## 2. INTRODUCTION

During the last years myopia control therapies have spawned a large interest among researchers and vision care professionals (Wolffsohn et al. 2016). Several methods have been proposed to slow myopia progression, among them, non-pharmacological treatments like orthokeratology and peripheral defocus modifying contact lenses (CLs) achieved very good outcomes (Huang et al. 2016), especially in patients with eso fixation disparity at near (Turnbull et al. 2016). The effect of such CLs is attributed to the induction of a myopic Relative Peripheral Refractive Error (RPRE) (Rodriguez-Vallejo et al. 2014; González-Méijome et al. 2016; Queirós et al. 2016; Walline 2016). Different designs of multifocal CLs were proposed to this aim (González-Méijome et al. 2016), and consequently, the amount and extension of the induced RPRE vary among lenses (Queirós et al. 2016).

In a previous paper (Rodriguez-Vallejo et al. 2014) we have proposed, and numerically validated, a new design of CLs for myopia control, named Fractal Contact Lenses (FCLs). However, the promising performance obtained with FCLs in model eyes has still not been validated in real eyes. Therefore, the main aim of this study is to assess the peripheral refraction induced by FCLs in myopic real eyes. To do that, RPRE was measured in a 2D matrix of discrete points of the retina; and this data set was represented as 2D power contour plot for the three components of the dioptric power vectors (M, $J_0$ and $J_{45}$). We show that this new representation of the RPRE, employed in this work for the first time in the literature, offers a more complete view of the lens performance than the usually used only along the horizontal field.

## 3. METHODS

*Contact Lenses*

FCLs prototypes were specially manufactured for this study. All lenses were made of Hioxifilcon A (Benz G5X p-GMA/HEMA)(Benz 2016), which has 1.401 of refractive index (hydrated and at 35º), using a precision lathe (Optoform 40, Sterling Ultra Precision, Largo, USA). A stock of 15 FCLs with treatment powers of +2.00 D was fabricated according to the design previously described (Rodriguez-Vallejo et al. 2014). The FCLs prototypes had a diameter of 14.50 mm, and central powers ranging from -0.50 D to -7.00 D in -0.50 D steps with two different base curves: 8.4 mm and 8.6 mm. Power profiles of the stock lenses were measured with the Nimo TR1504 (LAMBDA-X, Nivelles, Belgium)(Joannes et al. 2010) contact lens power mapper (version 4.2.6.0 r477). A custom function in MATLAB (R2013a; Mathworks, Inc., Natwick, MA) was developed in order to detect the power transition between therapeutic and compensation zones. Then, the true therapeutic powers for each lens were redefined as the difference of the mean power along the therapeutic zones and the mean power along the compensation zones.

*Subjects and Procedures*

Twenty-six subjects (mean age 23.77 ± 3.62 years) were recruited from students at the University of Valencia, Spain (18 females and 8 males). All underwent a complete eye exam including objective and subjective refraction and slit-lamp exploration. Inclusion criteria were myopic eyes ranging from -0.50 D to -7 D (mean -2.62 ± 1.59 D) and astigmatism equal or under -0.75 D with no ocular diseases, strabismus or amblyopia. Only right eyes were considered. The research adhered to the tenets of the Declaration of Helsinki, with the research approved by the University of Valencia and informed consent obtained from all participants.

Before fitting the FCLs, corneal topographies were taken for the naked eye with the Keratron Scout (Optikon 2000 SpA, Rome, Italy) until to obtain at least three of them with a reproducibility inside ±0.25D. Elevation data were exported in binary format (.XLB and .ZLB extension files) and a custom software was programmed in MATLAB in order to compute corneal asphericities, fitting elevation data to a conic function (Calossi 2007), at nasal and temporal sides from the normal vertex along the horizontal 0-180º, considering an extension of 4 mm from the vertex.

Subjects were fitted with the FCL that best matched one of the two possible base curves and the back vertex powers closer to the spectacle refraction. The behaviour of the lenses, movement and centration, were evaluated by the examiner twenty minutes after fitting. Then corneal topographies were taken again but with the patient wearing the best fitted FCL. The distance from the centre of the first therapeutic zone and the pupil entrance centre was measured with the caliper tool of the Keratron Scout software to obtain the FCL decentration.

*Peripheral Refractive Error*

Central and peripheral refraction were measured with an open-field autorefractor (Grand-Seiko WAM-5500, Grand-Seiko Co., Ltd., Hiroshima, Japan) in non-cyclopegic conditions. The environmental light was 150 lux, since at this condition the pupil diameter was higher enough to measure the peripheral refractive error. Tracking targets were distributed in a square grid from -30º (nasal retinal area) to 30º (temporal retinal area) and 15º superior to -15º inferior, as shown in Fig. 1. Fixation targets were 1 inch high contrast squares located on a wall at 2 meters from the eye. Measurements were taken with the eye rotation technique (Queirós et al. 2016).

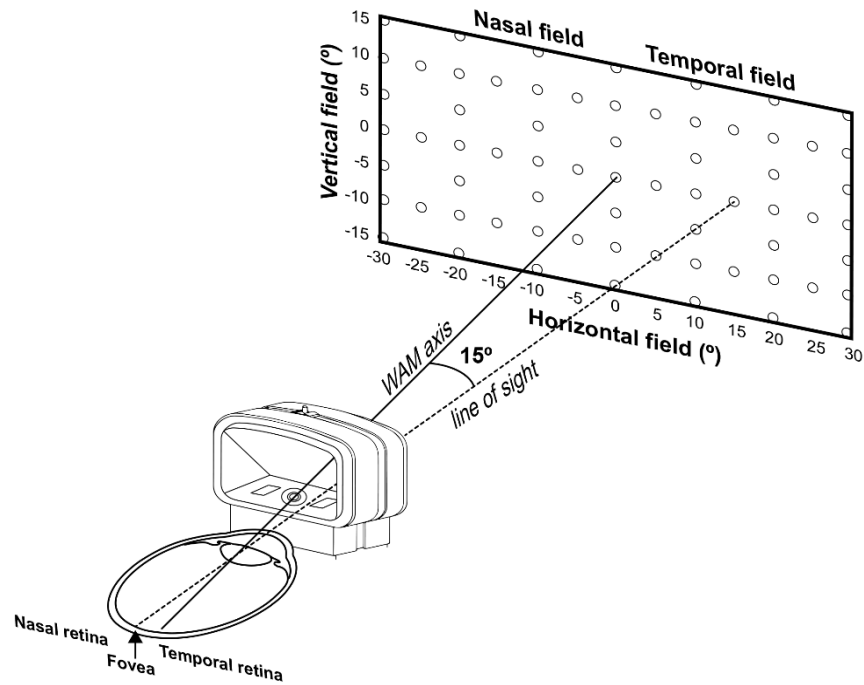

*Figure 1. Schematic representation of the measurement process with WAM-5500. With the right eye rotated 15 degrees looking at the temporal field target, the system is measuring the refractive error at the temporal retinal area. The circles over the wall represent the discrete points measured with the eye rotation.*

A MATLAB code was developed to obtain repeatable measures of sphere, cylinder and axis at each point of the field and to compute the mean values of vector components according to Fourier analysis (M, $J_0$ and $J_{45}$) (Thibos et al. 1997). The peripheral refractive error was measured without FCLs (baseline state) and wearing FCLs. Recorded data was used to compute, the tangential ($F_T = M + J_0$) and sagittal ($F_S = M − J_0$) power errors along the horizontal meridian (Queirós et al. 2016); and two dimensional RPRE contour maps for M, $J_0$ and $J_{45}$. These maps were calculated using custom software (Mathematica version 10; Wolfram Research, Inc., Oxfordshire, UK). A cubic interpolation was employed to represent contour lines of equal powers in steps of 0.12 D.

*Statistical Analysis*

Normal distributions were confirmed with the Shapiro-Wilk test. Paired t-tests were used to analyze the differences between the RPRE vector components with the FCLs and the nacked eye. Pearson correlation analyses were performed to determine the relationship between variables. Power analysis was performed using G Power version 3.1.9.2 (available at http://www.gpower.hhu.de/). The sample size in this study offered 88% statistical power at a 5% level to detect a difference in RPRE of 0.25 D with and without FCLs when the expected standard deviation (SD) of the mean difference was 0.44 D (obtained from previous measures). The data were managed using SPSS software version 20 (SPSS Inc., Chicago, IL, USA), and $p<0.05$ was considered to indicate significance.

## 4. RESULTS

*Contact Lenses: Power Profiles and Fitting*

The power profiles measured with NIMO resulted in a mean power of 1.32 ± 0.28 D at the therapeutic zones. Theoretical compensation power of the FCL prototypes was negatively correlated with the experimental therapeutic power (r = -0.786, p = 0.007).

Topological data revealed that contact lenses were decentered towards temporal cornea, ranging from 0.39 mm to 1.05 mm (mean 0.7 ± 0.19 mm) whereas mean vertical displacement was 0.00 ± 0.49 mm [ranging from 0.64 mm down to 1.38 mm up]. Considering the lens decentration in polar coordinates, the lenses were decentered a mean of 0.83 ± 0.27 mm at 185 ± 32 degrees. The mean value of the pupil entrance diameter was 3.67 ± 0.53 mm measured with the Keratron in the nacked eye.

*Horizontal Relative Peripheral Refractive Error*

Mean corneal asphericity from the corneal vertex to the 4 mm of semi-chord at temporal cornea was -0.07 ± 0.09 and -0.24 ± 0.18 for the same extension at nasal cornea in the nacked eyes. Significant correlations were found between the corneal asphericity and vector components of the RPRE (Table 1) and no correlations were found between the amount of lens decentering and the asphericity of the cornea along temporal and nasal sides.

*Table 1. Significant correlations between Relative Peripheral Refractive Error (RPRE) vector components and asphericity at the Temporal or Nasal semi-chord of the cornea from the normal vertex to 4 mm of radial position. Correlations for retinal areas evaluated and not represented in the table were not significant.*

| Retinal Area (°) | RPRE (D) Mean ± SD | Corneal side | Pearson r |
|---|---|---|---|
| **M** | | | |
| -25 (NR) | -0.22±0.47 | Temporal | -0.452, p=0.040 |
| -15 (NR) | -0.21±0.40 | Temporal | -0.526, p=0.014 |
| -10 (NR) | -0.27±0.29 | Temporal | -0.436, p=0.048 |
| **J0** | | | |
| +25 (TR) | -0.82±0.29 | Nasal | -0.572, p=0.007 |
| +20 (TR) | -0.56±0.22 | Nasal | -0.562, p=0.008 |
| +10 (TR) | -0.1±0.2 | Nasal | -0.505, p=0.019 |
| **J45** | | | |
| +30 (TR) | 0.11±0.25 | Nasal | -0.581, p=0.006 |
| +20 (TR) | 0.05±0.15 | Nasal | -0.465, p=0.033 |
| +10 (TR) | -0.01±0.09 | Nasal | -0.478, p=0.028 |
| +5 (TR) | 0.01±0.09 | Nasal | -0.467, p=0.033 |

NR=Nasal retina; TR= Temporal retina.

Fig. 2A shows the spherical equivalent (M) along the horizontal retinal area at the baseline state and wearing the FCLs. An increase of the myopic shift was found with the FCLs at the temporal retina from 10º to 30º (p<0.05). The $F_T$ myopic shift was increased after the FCLs fitting as it is shown is Fig. 2B with a peak located at 20º of the temporal retina, in the same way that the M component (Fig. 2A). $F_S$ was also increased myopically even though less markedly than $F_T$ but highly enough to move the sagittal foci to the front of the retina with the FCL (Fig. 2C).

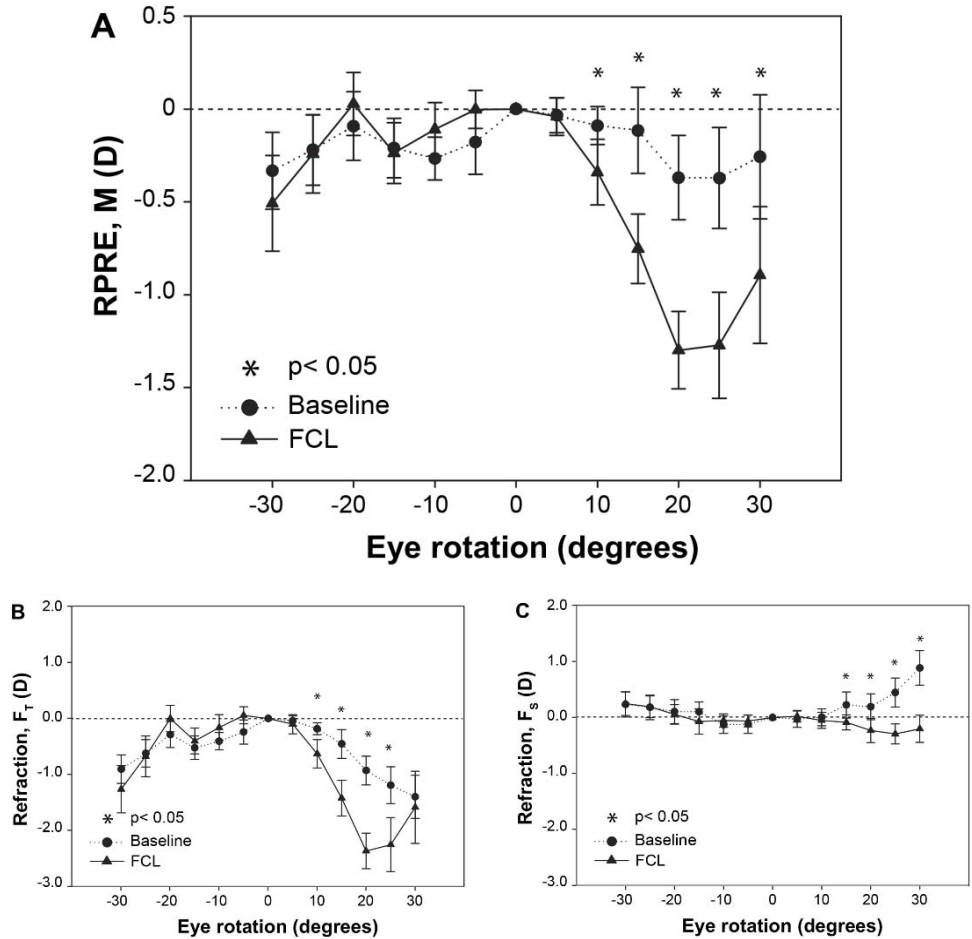

*Figure 2. (A) Refractive spherical equivalent in the nacked eye (Baseline) and with Fractal Contact Lenses (FCL). (B) Tangential and (C) Sagittal powers along the horizontal retina in the baseline state and with FCLs. Positive values of eye rotation correspond to the temporal retina and negative values to nasal retina. An asterisk over each eccentricity was represented to describe significant differences between baseline and with FCLs (p<0.05).*

*Two-Dimensional Relative Peripheral Refractive Error*

Baseline mean values of the RPRE for M, $J_0$ and $J_{45}$ are represented in Figs. 3A, 3D and 3G, respectively. Figs. 3B, 3E and 3H show the mean values for the same eyes wearing FCLs. The measured points are represented by means of circles over the difference maps (Figs. 3C, 3F and 3I). Crosses were drawn inside the circles for those positions where significant differences (p<0.05) between eyes with and without the FCLs were found.
Fig. 3C shows that for the spherical equivalent, the mean myopic shift induced by FCLs increases with the eccentricity and becomes significant ($p < 0.05$) at 10º in the temporal field. A clear oblique astigmatism was presented at extreme positions of the visual field in both situations (see Fig. 3G and Fig. 3H) with opposite signs for $J_{45}$ between temporal/nasal and inferior/superior. However, as can be seen in Fig. 3I, the oblique astigmatism induced by the FCLs was almost negligible along the horizontal and vertical central coordinates but significant in extreme positions.

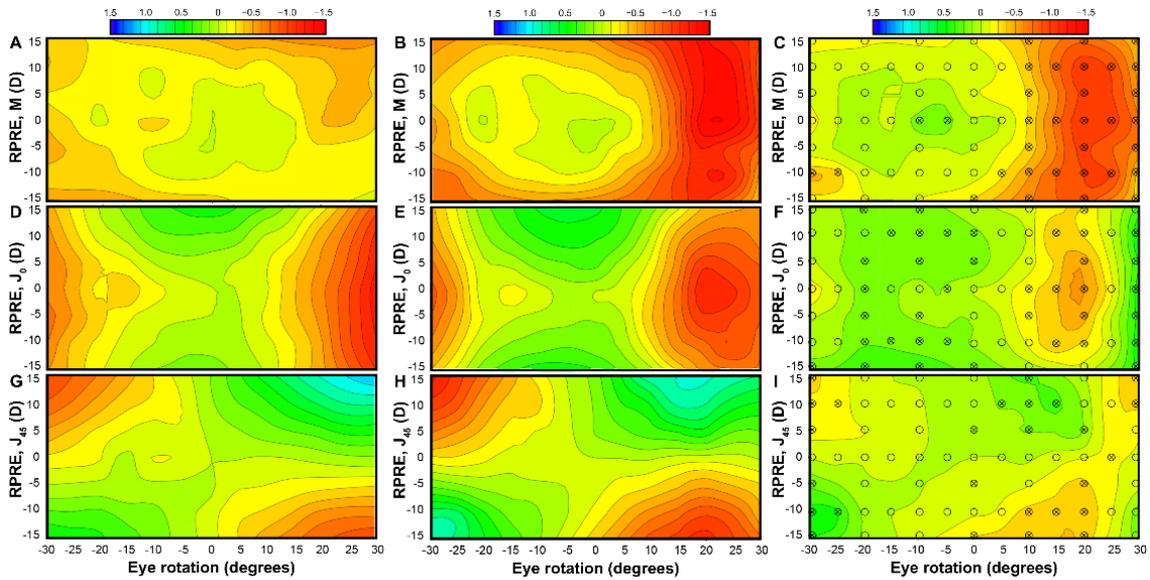

*Figure 3. Relative peripheral refractive error (RPRE) vector components M, J0 and J45 in the baseline state (A, D and G) and with FCLs (B, E and H). Differences for the three components between FCLs and baseline state are represented in C, F and I. Circles in the difference maps were used to describe the eye rotation for which direct measures were taken whereas crosses were used to indicate the rotations with significant differences (P<0.05). Color maps were obtained with cubic interpolations across the most nearest measured points.*

We found that myopic shift (M) induced by the FCLs at 25º and 30º along the temporal field decreased with the lens decentration through the temporal side of the cornea (r=0.46, p=0.022) and (r=0.47, p=0.018) respectively. These correlations were more marked when the displacement vector (considering x and y) was considered instead of the unique displacement along the horizontal, r = 0.50 (p = 0.013) at 25º and r = 0.54 (p = 0.006) at 30º.

## 5. DISCUSSION

Studies conducted on animals have established bases of the optical compensations to slow down myopia progression on humans. However, the mechanisms by which the visual system responds to the optical cues in the emmetropization process are still not clear. Experimental studies in animals found that refractive error in the peripheral retina can regulate the eye growth (Smith et al. 2005), and derived from these studies, relative peripheral hyperopia was suggested as a possible factor that could cause the progression of myopia (Smith 2012). Thus the goal in the current treatments, such as ortokeratology or multifocal contact lenses is to create a relative peripheral myopia to slowing down myopia progression (Huang et al. 2016). In this context, it is important to take into account that recent studies (Atchison et al. 2015; Hartwig et al. 2016), involving children and young adults, demonstrated that a relative peripheral hyperopia along the horizontal visual field does not predict development nor progression of myopia in these populations.

One of the possible explanations of these apparently contradictory results could be the lack of information in clinical studies about the behaviour of RPRE for the M, $J_0$ and $J_{45}$ on regions of the retina beyond the horizontal meridian. Related to this, findings obtained from animal studies suggest that the analysis of M component might not be enough to describe the emmetropization process (Queirós et al. 2016). Kee et. al found that emmetropization in infant monkeys can also vary with the imposition of cylindrical lenses by means of directing the process towards one of the two focal planes with independence of the astigmatism orientation (Kee et al. 2004). This is in a partial agreement with Chu et al. who also reported an eye expansion according to the amount of the cylinder in chicks, but in this case with dependence of the cylinder orientation (Chu and Kee 2015). Furthermore, the optical responses during the emmetropization process might vary depending on the stimulated retinal areas (Charman 2011) because of the neural anisotropy between central and peripheral retina (Zheleznyak et al. 2016). Therefore, to obtain information about RPRE for the three components M, $J_0$ and $J_{45}$ of the eye refraction seems to be of essential importance in understanding basic phenomena with important clinical consequences. In this sense, the main advantage of our approach from previous studies is that we have analysed the effect in the central and paracentral retina obtaining 2D information about the astigmatism at other positions besides the horizontal direction.

*Contact Lenses*

Some important issues should be taken into account in order to interpret our results and correlate them with the theoretical optical performance of the FCLs design (Rodriguez-Vallejo et al. 2014). First, the mean power at the treatment zones (+1.32 D) was lower than the theoretical power of +2.00 D. A negative correlation was also obtained between the compensation power of the prototypes and the power at the therapeutic zones, which means that high power minus lenses had less power in the therapeutic zones than lower power FCLs. This suggests that the manufacturing process should be optimized because the lenses would have less therapeutic power in myopes who have higher peripheral relative hyperopia (Atchison 2006). The FCLs fitted to the subjects also presented a mean decentration of 0.83 mm. The theoretical performance of the FCL was computed in a previous work (Rodriguez-Vallejo et al. 2014) for a mean decentration of 0.7 mm obtaining a peak myopic shift of around -2 D at 30º. Our results in this study have shown a peak around -1.3 D (the same value of the mean treatment power of the prototypes) at 20º-25º (Fig. 2A). Thus, in comparison to the ray tracing analysis (Rodriguez-Vallejo et al. 2014), the displacement of the peak can be explained by the decentration of the lens whereas the undercorrection the periphery, is justified by the lower experimental power at the treatment zones of the prototypes.

*Horizontal Relative Peripheral Refractive Error*

For the nacked eye we found that corneal asphericity along temporal and nasal semichords in the horizontal meridian was negatively correlated with the M; but only for temporal cornea and nasal retina whereas for $J_{45}$ and $J_0$ the negative correlations where found between nasal cornea and temporal retina (Table 1). This is in agreement with theoretical models, which assert that: as more positive is the asphericity (Q), more myopic is the peripheral refraction induced for M and $J_0$ (Atchison 2006; He 2014). Therefore, corneal asphericity has a correlation with RPRE and this fact should be considered by clinicians, who usually cannot measure the RPRE, because as less prolate (more positive Q) is the corneal semi-meridian as more myopic is the RPRE and more prolate corneas might be expected to progress into more myopia (Horner et al. 2000). We also found a trend for the $J_{45}$ to become more positive with increasing the eccentricity in the temporal retinal area and more negative in the nasal area, whereas $J_0$ become more negative in both sides of the retina. These results agree the previous reported in other studies that measure peripheral refraction in myopes (Atchison et al. 2006; Davies and Mallen 2009; Radhakrishnan et al. 2013).

At the baseline state, the sample showed a relative peripheral myopia lower than -0.50D for M, which agrees the previous reported studies considering that in our study the mean value of the central refraction was -2.62 D (Atchison 2006). Sagittal power was hyperopic along the temporal retina (Fig. 2C), but become myopic with the FCLs. This is an advantage of FCLs for myopia control over the ortokeratology because in ortokeratology, especially in low myopes, it cannot be guaranteed that the sagittal power became myopic (González-Méijome et al. 2015), whereas with FCLs the power at treatment zones can be modified to achieve higher values of RPRE (Queirós et al. 2016).

*Two-dimensional Relative Peripheral Refractive Error*

The two dimensional representation of the RPRE offered us further information about what happens in a wide area of the retina, especially considering the lens decentration. The mean M showed a significant increase of the myopic shift in the temporal retina that was almost uniform along the explored vertical field (Fig. 3C) which means that only representing the horizontal section such as in Fig. 2A would be enough to know the change in RPRE for M with this kind of FCLs. Furthermore, although the effect of the FCLs on the spherical equivalent M is almost uniform around the centre it was not symmetric from nasal to temporal retina, which is explained by the lens decentration to the temporal cornea. Note that the same asymmetry was obtained for $J_0$ and $J_{45}$. However, the changes in the $J_0$ (Fig. 3F) and $J_{45}$ (Fig. 3I) along the vertical were not uniform as happened with M, that suggests an advantage of the two-dimensional representation for increasing the understanding about the changes in the astigmatism components due to the lens. Particularly, we found that the sign of the $J_{45}$ component changed in extreme positions of the superior and inferior retina (Fig. 3G), which could be explained by a modification in the orientation of the main focal points due to the astigmatism originated by the oblique incidence of the light over the optical structures. This finding is also in agreement with other authors (Ehsaei et al. 2011) and it has been demonstrated by ray tracing models (Rojo et al. 2015). The components $J_0$ and $J_{45}$ were more myopic with the FCLs (Figs. 3F and 3I), and the peaks of maximum induction power were located near to the same area of the peak of M (Fig. 3C).

*Conclusions*

Despite peripheral defocus modifying CLs proved their effectiveness in slowing down myopia progression (Huang et al. 2016), the optical theories for which these lenses are founded remain controversial (Radhakrishnan et al. 2013; Atchison et al. 2015; Hartwig et al. 2016). New studies on animals with dual focus lenses are increasing the knowledge about the optical signals that modulate the eye growing. For instance, it was recently found that in the presence of more than one foci, the refractive development was dominated by the more anterior image plane (Arumugam et al. 2014) and also, that the refractive development is dominated by the relative myopic defocus even though the therapeutic area is reduced to one-fifth the total area (Arumugam et al. 2016). Furthermore, the peripheral retinal area might respond in a different way to optical defocus than the central retina due to the neural anisotropy in the paracentral retina (Zheleznyak et al. 2016). While appear new findings about optical signals that modulate eye growing in research literature, the complete representation of the changes produced by peripheral defocus modifying CLs can help to understand the nature of the optical signals produced by these lenses.

In summary, a new two-dimensional representation of the RPRE was proposed. This representation was employed for validating the performance of the FCLs (Rodriguez-Vallejo et al. 2014). Very good agreement between the theoretical prediction and the experimental results was obtained. In fact, differences are explained considering both the imperfections in the manufacturing process of the prototypes, and the lens decentration.

## 6. ACKNOWLEDGEMENTS

This study was supported by the Ministerio de Economía y Competitividad and FEDER (Grant DPI2015-71256-R), and by the Generalitat Valenciana (Grant PROMETEOII-2014-072), Spain.

## 7. CONFLICT OF INTEREST

The author(s) have made the following disclosure(s):

Karina Georgieva: None.

Vicente Ferrando: None.

M. Rodríguez-Vallejo, J. A. Monsoriu and W. D. Furlan inventors (P) ES Patent P201330862, relating to contact lens design: assigned to Universidad Politécnica de Valencia and Universitat de Valéncia.